\documentclass{sig-alternate-05-2015}

\usepackage[T1]{fontenc}

\usepackage{algpseudocode}
\usepackage{algorithm}
\usepackage{fixltx2e}
\MakeRobust{\Call}

\usepackage{array}

\usepackage{hyperref}

\usepackage{flushend}

\usepackage{xcolor}
\usepackage{changebar}

\usepackage{graphicx}
\graphicspath{ {./figures/} }

\usepackage{tabularx}

\title{Fast Online k-nn Graph Building}

\numberofauthors{3}
\author{
    \alignauthor Thibault Debatty\\
           \affaddr{Royal Military Academy}\\
           \affaddr{Brussels, Belgium}\\
           \texttt{thibault.debatty@rma.ac.be}
    \alignauthor Pietro Michiardi\\
           \affaddr{EURECOM}\\
           \affaddr{Campus SophiaTech, France}\\
           \texttt{pietro.michiardi@eurecom.fr}
    \alignauthor Wim Mees\\
           \affaddr{Royal Military Academy}\\
           \affaddr{Brussels, Belgium}\\
           \texttt{wim.mees@rma.ac.be}
}

\setcopyright{acmcopyright}
\conferenceinfo{KDD '16}{August 13--17, 2016, San Francisco, CA, USA}

\DeclareMathOperator{\similarity}{similarity}

\begin{document}

\maketitle

\begin{abstract}

In this paper we propose an online approximate $k$-nn graph building algorithm, which is able to quickly update a $k$-nn graph using a flow of data points. One very important step of the algorithm consists in using the current distributed graph to search for the neighbors of a new node. Hence we also propose a distributed partitioning method based on balanced $k$-medoids clustering, that we use to optimize the distributed search process. Finally, we present the improved sequential search procedure that is used inside each partition.

We also perform an experimental evaluation of the different algorithms, where we study the influence of the parameters and compare the result of our algorithms to existing state of the art. This experimental evaluation confirms that the fast online $k$-nn graph building algorithm produces a graph that is highly similar to the graph produced by an offline exhaustive algorithm, while it requires less similarity computations.

\end{abstract}

\section{Introduction}

A $k$-nn graph is a data structure where each element (called node or vertex) has a link (an edge) to the $k$ most similar elements of the dataset. Building a $k$-nn graph is a time consuming operation, as it requires the computation of $O(n^2)$ similarities, where $n$ is the number of elements in the dataset. At the opposite, analyzing a $k$-nn graph is usually a fast operation. Therefore, $k$-nn graphs are often used for interactive data analytics, like clustering for example.

In the general case, building a $k$-nn graph requires the computation of $n \cdot (n-1) / 2$ similarities. If the similarity used is a metric, the triangle inequality can be used to reduce the number of computations. In anyway, building an exact $k$-nn graph remains a computationally heavy process.

Therefore, research mainly focuses on building approximate $k$-nn graphs, where each node has edges to the $k$ most similar nodes with a high probability. These algorithms can be grouped in two categories: 1) algorithms that partition the dataspace to reduce the number of similarities to compute and 2) algorithms that use the hill climbing principle to iteratively improve the graph.

In the same way, building an online $k$-nn graph from a flow of data points requires, for each new node: 1) to compute the edges of the new node and 2) to update the edges of existing nodes. Therefore, each new data point requires to compute the similarity between the new point and every node in the existing graph, which is very computation intensive. However, current approaches are still slow, and better alternatives that can achieve higher speedups w.r.t. a naïve approach are truly desirable.

Therefore, in this paper we propose an online approximate $k$-nn graph building algorithm, which is able to quickly update a $k$-nn graph using a flow of data points. To the best of our knowledge, this is the first algorithm of this kind. We present a sequential and a distributed version of the algorithm. Moreover, our algorithm is independent of the similarity measure used to build or query the graph.

The distributed algorithm starts with an initial $k$-nn graph that is first partitioned using a distributed balanced k-medoids algorithm. This partitioning is used to improve distributed graph based search. Indeed, the algorithm has two main steps to add a new point to the graph: 1) use the distributed graph to search the $k$ nearest neighbors of the new pointand 2) update the graph: these neighbors are used as starting points to search existing nodes for which the new point is now a nearest neighbor. To search the nearest neighbors, inside each partition the algorithm uses an fast sequential graph based nearest neighbor search algorithm.

The rest of this paper is organized as follows. In Section~\ref{sec:related} we present existing graph building algorithms, graph based search algorithms and graph partitioning algorithms. In Section~\ref{ignns} we present our improved graph based search method. In Section~\ref{sec:seq-online} we present the sequential version of the online $k$-nn graph building algorithm. We then head over to the distributed algorithms. In Section~\ref{sec:distributed-search} we explain the distributed graph based search method, which relies on the k-medoids based partitioning method that we present in Section~\ref{sec:distributed-kmedoids}. In Section~\ref{sec:distributed-online} we present the distributed online graph building algorithm. In Section~\ref{sec:evaluation} we perform an experimental evaluation, where we perform a parameter study of the algorithm and compare it against existing state of the art. Finally, in Section~\ref{sec:conclusion}, we present our conclusions and propositions for future work.

\section{Related work}
\label{sec:related}

We present here related work in the domain of $k$-nn graph building. One important step in our algorithm consists in searching the nearest neighbors of the new data point using the existing graph. Therefore we also present existing graph based nearest neighbor search algorithms. Finally, searching the graph in a distributed fashion requires a specific partitioning of the graph. Hence we present here existing graph partitioning algorithms.

\subsection{Graph building}

Different approaches exist to build a $k$-nn graph. Some of them tolerate incorrect edges to speedup the building process and produce an approximate graph, while others produce an exact graph. In both cases, these building algorithms are closely related to nearest neighbor search algorithms.

The naive method, also called linear search, consists in computing the distance between the query point and every other point in the set, keeping track of the ``best node so far'' (or $k$ ``best nodes so far''). Similarly, the most naive way to build a complete $k$-nn graph is is to use brute force to compute all pairwise similarities. Then, for each node, the algorithm keeps only the $k$ edges with the highest similarity. This method has a computational cost of $O(n^2)$ and is thus very slow, even implemented in parallel.

Another approach is to use some kind of index to speedup nearest neighbors search. These techniques usually rely on the branch and bound algorithm, and the index is used to partition the data space. For example, a $k$-$d$ tree, that recursively partitions the space into equally sized sub-spaces, can be used to speedup neighbor search~\cite{Moore1991}. R-trees~\cite{Guttman:1984:RDI:602259.602266} can also be used for euclidean spaces. In the case of generic metric spaces, vantage-point trees~\cite{Fu:2000:DVI:765229.765232}, also known as metric trees~\cite{Uhlmann1991}, and BK-trees can be used. But these approaches are hard to implement in parallel on a shared nothing architecture like MapReduce (MR) or Spark. In \cite{Connor2009} for example, the authors present a distributed $k$-nn graph building algorithm, but use a shared memory architecture to store a kd-tree based index.

Some nearest neighbors search algorithms use Locality-Sensitive Hashing (LSH), like \cite{Rajaraman2010}, to hash the input items so that similar items are mapped to the same buckets with a high probability. As opposed to conventional hash functions, such as those used in cryptography, the goal of LSH is to maximize the probability of collision between similar items. Various authors also propose algorithms relying on LSH the build $k$-nn graphs. In~\cite{Zhang2013}, the authors use LSH to divide the dataset into small groups. Then, inside these small groups, the algorithm builds the $k$-nn graph. As groups are not overlapping, the constructed graph is a union of multiple isolated small graphs. To build the final graph, and improve the approximation quality, the division is repeated several times to generate multiple approximate graphs, which are combined to produce the final graph. They also show experimentally that their algorithm is much faster than existing algorithms, for similar quality of the built graph. However, LSH approaches can only be used for some similarity measures: $l_p$, Mahalanobis distance, kernel similarity, and $\chi^2$ distance.

A different and versatile algorithm to efficiently compute an approximate $k$-nn graph is described in \cite{Dong2011}. The algorithm, called nn-Descent, starts by creating edges between random nodes. Then, for each node, it computes the similarity between all neighbors of the current neighbors, to find better edges. The algorithm iterates until it cannot find better edges. The main advantage of this algorithm is that it works with any similarity measure.

\subsection{Graph based nn-search}

The nearest-neighbor search problem (NN search) is formally defined as follows: given a set $S$ of points in a space $M$ and a so-called query point $q \in M$, find the closest point in $S$ to $q$, according to some similarity metric. The $k$-nn search is a direct generalization of this problem, where we need to find the $k$ closest points.

A lot of algorithms exist to find the $k$ nearest neighbors of a point. They are generally very similar to those used to build a $k$-nn graph. However, only a few of them rely on an existing $k$-nn graph to find the nearest neighbors of a query point.

In \cite{Hajebi2011}, Hajebi et al. proposed a new sequential approximate NN search algorithm that relies on $k$-nn graphs. The algorithm, called Graph Nearest Neighbor Search (GNNS), works by selecting initial nodes at random. For each node, the algorithm computes the similarity between query point and every neighbor. The most similar neighbors are selected, and the algorithm iterates until a depth of search $d$ is reached. It is thus a ``hill climbing'' algorithm. The most promising nodes are searched first, using the similarity between the query point and the node as a heuristic.

It was tested against different datasets. Without taking graph building phase in account, the search algorithm achieved a speedup of up to 80 over linear search, and a speedup of two over randomized KD-tree.

Dong, the coauthor of the paper on nn-descent~\cite{Dong2011}, also created a software called KGraph~\cite{kgraph} which is able to search the nearest neighbors of a query point using a precomputed $k$-nn graph. However, the search algorithm used by the program was never published.

\subsection{Graph partitioning}

The classical definition of graph partitioning consists in splitting the graph data between partitions, minimizing the number cross partition edges, while keeping the number of nodes in every partition approximately even.

Multiple algorithms exist to perform graph partitioning. In \cite{Rahimian2013}, the authors proposed a distributed iterative algorithm that iteratively swaps the partition of two nodes to minimize the number of cuts. The algorithm is heavily based on MPI and requires a lot of communication between all nodes of the graph. In \cite{Carlini2014}, the authors proposed and tested a Bulk Synchronous Parallel (BSP) version of the algorithm which makes it suitable for shared nothing architectures like Apache Spark. In \cite{Kliot2012}, the authors proposed a streaming algorithm, that requires a single iteration to partition the graph. They experimentally compared various heuristics to assign nodes to a partition. They found the best performing heuristic was linear weighted deterministic greedy. This one assigns each node to the partition where it has the most edges, weighted by a linear penalty function based on the capacity of the partition.

As we show in Section~\ref{sec:distributed-search}, to improve distributed graph based search, the partitioning scheme should minimize the number steps between any two nodes in the partition. In this case the partitioning becomes a k-medoids clustering problem. It is a variation of $k$-means clustering, where the centers are datapoints. It also minimizes the sum of pairwise distances, while $k$-means minimizes the sum of squared Euclidean distances. Just like k-means clustering, various algorithms were proposed in the literature to perform $k$-medoids clustering, like Partitioning Around Medoids (PAM) \cite{Theodoridis:2006:PRT:1200914}. To the best of our knowledge, the most efficient algorithm for performing $k$-medoids clustering is currently the Voronoi iteration method proposed in \cite{Park2009}, which is very similar to the classical Lloyd's algorithm used to compute k-means. Until now, no balanced version of k-medoids was published.

However, when it comes to k-means, few balanced versions exist. In \cite{Malinen2014}, the authors proposed a method that has a complexity $O(n^3)$, which makes it too complex for large graphs. In \cite{Banerjee2002}, the authors proposed the Frequency Sensitive Competitive Learning (FSCL) method, where the distance between a point and a centroid is multiplied by the number of points already assigned to this centroid. Bigger clusters are therefore less likely to win additional points. In \cite{Althoff2008}, the authors used FSCL with additive bias instead of multiplicative bias. However, both methods offer no guarantee on the final number of points in each partition, and experimental results have shown the resulting partitioning is often largely imbalanced.

\section{Improved graph based nn-search}
\label{ignns}

An important step of our online graph building algorithm consists in using the existing graph to search the $k$ nearest neighbors of the new data points. Hence we present here a new fast approximate nearest neighbor search algorithm, which is able to find the nearest neighbors of a point with a high probability, while computing only a few similarities. The search algorithm has the additional advantages that is can be used with any similarity measure, and that the only required parameter is the desired speedup compared to exhaustive search, which makes it very easy to tune.

To search the nearest neighbors of a query point $q$ we use the same hill climbing approach as the one used by the Graph NN Search (GNNS) algorithm presented in~\cite{Hajebi2011}: the algorithm selects a random node $n$ from the graph, computes the similarity between $q$ and every neighbor of $n$, and iterates with the most similar neighbor. While iterating, it keeps a set of the most similar points to $q$. When a local maximum is reached, the algorithm restarts with another random node. This hill climbing approach allows the algorithm to work with any similarity measure, metric or not. This search algorithm is thus an approximate algorithm, as it does not necessarily find the most similar node in the graph. It does however find the nearest neighbor with a high probability, while analyzing only a fraction of the nodes in the graph.

However, we introduce two major improvements, hence we call this search algorithm improved GNNS (iGNNS). These improvements are actually additional approximations, which allow iGNNS to further reduce the number of similarities to compute. Hence, for the same probability of finding the nearest neighbors, iGNNS provides an increased speedup compared the original GNNS.

The first improvement relies on the observation that by the definition of a $k$-nn graph, each node only has edges to other very similar nodes. Hence the increase of similarity at each iteration of the search can be very small. As a consequence, the number of iterations $i$ (the number of nodes to analyze) before finding the nearest neighbors of a query point can be very large. In the worst configuration of the graph, $i = n / k$. This requires to compute a lot of similarities, $i \cdot k$. To avoid this situation, the randomly chosen starting node $r$ is skipped if it is situated too far from the query point.

Formally, we keep track of the similarity of the most similar neighbor found so far $s_{\text{max}}$, and we introduce an expansion coefficient $e > 1$. As stated above, when a local maximum is reached, the algorithm restarts with another random node $r$. We immediately discard $r$ and select a new random node if $\similarity (\mathit{query}, r) < s_{\text{max}} / e $. In this way, we avoid analyzing a potentially very long chain of $i$ nodes before reaching the neighborhood of the query point, which would be computationally very expensive ($i \cdot k$). Instead, we focus on exploring the vicinity of the query point (the nodes for which similarity with query point is at least $s_{\text{max}} / e$).

Secondly, we further reduce the number of similarities computed at each iteration. In the original hill climbing approach, at each iteration the algorithm computes the similarity between the query point and every $k$ neighbors of the current node under test. Instead, we eagerly iterate using the first neighbor that provides an increase in similarity compared to the currently analyzed node. The improvement provided can be calculated for a euclidean space of $d$ dimensions and uniformly randomly distributed data points. In such a space, observe that for any node, on average only $e_h = k/2^d$ edges lead to a node with higher similarity, and $e_l = k - k/2^d$ edges lead to a node with lower similarity. As the improved algorithm iterates as soon as a node with higher similarity is found, the expected number of similarities to compute for each analyzed node is:

$$e_s = \frac{k - k/2^d}{1 + k/2^d}$$

At the opposite, the original hill climbing approach requires to compute $k$ similarities for each analyzed node. Hence this results in a speedup of $k / e_s$ compared to the original hill climbing approach. The resulting speedup for some values of $k$ and $d$ is shown in Table~\ref{table:search-eager}, which shows a substantial speedup can be achieved in some cases. The drawback is that in some cases the algorithm might pick a neighbor that improves the similarity, but without maximizing it (there was another neighbor which is more similar to the query point). However, a node has edges only to other highly similar nodes, hence those neighbors are also similar to each other. The difference of similarity improvement when choosing a sub optimal neighbor remains thus limited, and globally the efficiency of the algorithm increases.

Finally, the original GNNS algorithm stops when a fixed number of restarts has been reached, while iGNNS stops after a given number of similarities have been computed. This makes it easy to tune the algorithm, as the only required parameter is the speedup compared to exhaustive search. The computational cost of the algorithm is thus $n / \textit{speedup}$, where $n$ is the size of the graph.

\begin{table}[]
\centering
\caption{Speedup compared to classical hill-climbing search achieved by iterating as soon as a node with higher similarity is found, for various values of $k$ and $d$ (dimensionality).}
\label{table:search-eager}
\begin{tabular}{l l l l l}
\hline
$k$      & 4    & 10   & 10   & 10      \\
$d$      & 2    & 2    & 3    & 4       \\ \hline
$e_h$    & 1    & 2.5  & 1.25 & 0.625  \\
$e_l$    & 3    & 7.5  & 8.75 & 9.375  \\ \hline
$e_s$    & 1.5  & 2.14 & 3.88 & 5.77    \\
speedup  & 2.66 & 4.66 & 2.57 & 1.73    \\ \hline
\end{tabular}
\end{table}

\section{Sequential fast online k-nn\\graph building}
\label{sec:seq-online}

Now that we have presented the improved search algorithm, we present the sequential version of the fast online $k$-nn graph building algorithm. As shown in Algorithm~\ref{algo:seq-online-build}, the algorithm works in two main steps to add a new node to the graph: 1) search: it uses the existing graph to find the $k$ nearest neighbors of the new point and 2) update: it uses these neighbors as starting points to search existing nodes for which the new point is now a nearest neighbor.

\begin{algorithm}
  \caption{Sequential online graph build}
  \label{algo:seq-online-build}
  \begin{algorithmic}
    \State Inputs:
    \State $\textit{new}$: the new node 
    \State $\textit{graph}$: the existing graph 
    \State
    \State $\textit{neighborlist}$ = iGNNS($\textit{graph}$, $\textit{new}$, $k$) \Comment{Search}
    \State
    \State Update(\textit{graph}, \textit{new}, \textit{neighborlist}) \Comment{Update}
  \end{algorithmic}
\end{algorithm}

The update procedure is listed in Algorithm~\ref{algo:update}. It starts with the discovered neighbors of the new node, and iteratively explores the neighbors of neighbors, up to a fixed depth.

\begin{algorithm}
  \caption{Update}
  \label{algo:update}
  \begin{algorithmic}
    \State Inputs:
    \State \textit{graph}: the current graph
    \State \textit{new}: the new node to add in the graph
    \State \textit{neighborlist}: the neighbors of \textit{new}
    \State DEPTH: the depth of exploration
    \State
    \State Let $Q$ the list of nodes to analyze 
    \State Let $Q_{\text{next}}$ the list of nodes to analyze at next depth
    \State $Q$.addAll($\textit{neighborlist}$)
    \For{d in 1..DEPTH}
      \While{$Q$.hasNodes()}
        \State $\textit{node}$ = $Q$.pop()
        \State $Q_{\text{next}}$.addAll(neighbors of \textit{node})
        \State compute similarity(new, node)
        \State if needed, add \textit{new} to the neighborlist of \textit{node}
      \EndWhile
      \State $Q$.addAll($Q_{\text{next}}$)
      \State $Q_{\text{next}}$.empty()
    \EndFor
  \end{algorithmic}
\end{algorithm}

In this algorithm, the update step requires a maximum of $k^{\text{DEPTH}}$ similarity computations. The total computation cost of Algorithm~\ref{algo:seq-online-build} is thus

$$O(\frac{n}{\textit{speedup}} + k^{\text{DEPTH}})$$

To reduce the space requirement of the graph, $k$ is generally kept small. A value of 10 is very often seen. With this value, experimental evaluation has shown that a depth of two is sufficient to update the graph. The resulting number of similarity computations (100) is thus small compared to the size of the graphs targeted by this online building algorithm. The computation cost of the algorithm will thus be dominated by the search step, hence the need for a very efficient algorithm, such as iGNNS.

\section{Distributed graph based\\nn-search}
\label{sec:distributed-search}

We now head over to the distributed version of the algorithm. Just like the sequential version, the first and most important step of the distributed online $k$-nn graph building consists in searching the nearest neighbors of a new point using the existing graph. We hence start by presenting the distributed graph based nn-search algorithm.

The usual representation of a $k$-nn graph consists of adjacency lists (a list of the $k$ nearest neighbors of a single node). In a distributed environment, the $n$ adjacency lists are usually randomly distributed between $p$ partitions, which are processed by the different compute units. The number of partitions $p$ is usually chosen by the administrator, as a function of the number of compute units.

The procedure to perform $k$-nn search, presented in Algorithm~\ref{algo:distributed-search}, is actually very simple: the $p$ partitions are searched independently using iGNNS, then the $k \cdot p$ nearest neighbor candidates are filtered to keep $k$ most similar to the query point. Like the sequential algorithm, the distributed search algorithm requires to compute $n / \textit{speedup}$ similarities. The data exchanged is very limited: the compute nodes send the $k \cdot p$ candidate neighbors to the master.

\begin{algorithm}
  \caption{Distributed search}
  \label{algo:distributed-search}
  \begin{algorithmic}
    \State In parallel:
    \State In each subgraph, search $k$ nearest neighbor candidates
    \State using iGNNS
    \State
    \State Keep $k$ most similar nodes
  \end{algorithmic}
\end{algorithm}

As iGNNS travels the graph following edges, the graph must be partitioned in a very specific way to maximize the probability of reaching the nearest neighbors of the query point. The partitioning scheme to split the $k$-nn graph into subgraphs is actually a clustering algorithm, where the number of clusters is known ($p$). To maximize the probability of finding the most similar nearest neighbors, the partitioning scheme should fulfill two conditions: 1) the distance (the number of edges) between two nodes in the same sub-graph should be as low as possible, to maximize the probability of quickly finding ``good'' candidates and 2) as the sequential search algorithm will compute the same number similarities in each subgraph, the number nodes in each sub-graph should be similar. This last condition is also mandatory to balance the work load between compute nodes.

The first condition corresponds to the definition of $k$-medoids clustering, a variation of $k$-means clustering where the centers are data points. It also minimizes the sum of pairwise distances, while $k$-means minimizes the sum of squared Euclidean distances. To the best of our knowledge, a distributed balanced k-medoid algorithm was never proposed in the literature. Therefore we present our own in the next section.

\section{Distributed balanced\\k-medoids clustering}
\label{sec:distributed-kmedoids}

We present here the algorithm used to partition an existing graph for optimizing distributed graph based search [Algorithm~\ref{algo:distributed-search}]. The algorithm is actually a distributed balanced k-medoids clustering algorithm. To achieve a balanced distribution of points between the $k$ clusters (not to confuse with the $k$ edges per node of a $k$-nn graph), we use a linear weighted deterministic greedy heuristic: a node is assigned to the most similar medoid, weighted by a penalty function based on the capacity of the cluster which penalizes large clusters:

$$w(t, m) = 1 - \frac{|C_t(m)|}{\textit{capacity}} $$

where $C_t(m)$ is the cluster corresponding to medoid $m$ at time $t$ and $\textit{capacity}$ is the maximum size of each cluster. Generally a small imbalance can be tolerated between clusters, hence the capacity of clusters is computed using an imbalance factor ($\textit{imbalance} \geqslant 1$), where a perfectly balanced clustering can be achieved using $\textit{imbalance} = 1$:

$$\textit{capacity} = \frac{n \cdot \textit{imbalance}}{k}$$

To execute the algorithm in a distributed, shared nothing environment, we first randomly distribute the input dataset between the $c$ compute nodes. Hence the capacity constraint in each compute node becomes

$$\textit{capacity} = \frac{n \cdot \textit{imbalance}}{k \cdot c}$$

Then, during the update step, we compute the new medoids. The complete clustering algorithm is presented in Algorithm~\ref{algo-distributed-balanced-kmedoids}.

\begin{algorithm}
  \caption{Distributed balanced k-medoids}
  \label{algo-distributed-balanced-kmedoids}
  \begin{algorithmic}
    \State Input: dataset $d$
    \State
    \State Randomly pick $k$ initial medoids 
    \State Randomly distribute $d$ between compute nodes
    \Loop
      \State   In parallel: \Comment{Assign}
      \State   Let $d' \leftarrow$ copy $d$, assigning each node $n$ 
      \State   to the medoid $m$ that maximizes
      \State   $\text{similarity}(m, n) \cdot w(t, m)$
      \State
      \State   Shuffle $d'$ using medoid as key
      \State
      \State   In parallel: \Comment{Update}
      \State   use $d'$ to compute new medoids
    \EndLoop
    
  \end{algorithmic}
\end{algorithm}

The algorithm requires a copy of the whole dataset, and hence has a space requirement of $O(2n)$. As the complete dataset is distributed between compute nodes at each iteration, the communication cost is $O(i \cdot n)$ where $i$ is the number of iterations. The computation cost of a single assign step is $O(k \cdot n)$. The parallelism of the update step is maximum $k$. If the clustering of the data points is perfectly balanced, during the update step the size of each cluster is $n / k$. For each cluster, computing the new medoid requires to compute every pairwise similarities:

$$ \frac{n}{k} \cdot \frac{\frac{n}{k} - 1}{2} \approx O(\frac{n^2}{k^2})$$

The lower bound on the total computational cost for computing the $k$ new medoids is thus 

$$O(\frac{n^2}{k})$$

The total computational cost of the algorithm is

$$O(i \cdot (kn + \frac{n^2}{k}))$$

and as $k \ll n$, the computational cost of the distributed balanced $k$-medoids clustering algorithm is

$$O(i \cdot \frac{n^2}{k})$$

In our case, we wish to cluster the graph in order to optimize distributed search. Hence the distance used to compute the new medoids is the length of the shortest path in the graph between two nodes. These shortest paths are computed using Dijkstra algorithm. Although the update step requires to compute $O(n^2/k)$ distances, these are very fast to compute. The computation cost is hence dominated by the assign step.

Indeed, during the assign step, computing the shortest path from a node to all medoids requires to travel the complete graph. It requires thus access to all adjacency lists, which is impossible in a distributed, shared-nothing infrastructure. Therefore, instead of computing the shortest path to every medoid, we use the similarity between the node and every medoid as a heuristic. An intuitive example is to consider the simple case of a uniform distribution over an euclidean space. In such a space, all edges have the same length. Hence the most similar medoid, measured as the shortest path from the node to the medoid, is also the most similar medoid, measured using the euclidean distance. The computation cost to partition the graph to optimize the graph for distributed search is thus actually $ O(ikn) $.

\section{Distributed online k-nn graph building}
\label{sec:distributed-online}

We present here the distributed version of the online $k$-nn graph building algorithm. It starts with an initial $k$-nn graph that is first partitioned using the distributed balanced k-medoids algorithm presented in Section~\ref{sec:distributed-kmedoids}.

Just like the sequential algorithm, it has two main steps to add a new point to the graph: 1) use the current graph to search the $k$ nearest neighbors of the new point, using the distributed algorithm presented in Section~\ref{sec:distributed-search} and 2) update the graph using the procedure presented in Algorithm~\ref{algo:update}. In addition, the new node is assigned to the compute node corresponding the the most similar medoid.

Finally, the medoids may be recomputed once a given number of new nodes have been added to the graph. This is actually not mandatory, and depends on the dataset: if the characteristics of the dataset are fixed over time, adding new nodes will not induce a displacement of the medoids. Otherwise, the medoids update rate should be consistent with the expected rate of change of the dataset. The automatic estimation of the update rate is left as as future work. The complete procedure used to add a new node to the graph is shown in Algorithm~\ref{alg:distributed-online}.

\begin{algorithm}
  \caption{Distributed online $k$-nn graph building: add a node to the graph}
  \label{alg:distributed-online}
  \begin{algorithmic}
    \State Inputs:
    \State $\textit{graph}$: current graph
    \State $\textit{node}$: a new node
    \State
    \State In parallel: \Comment{Search with Algorithm~\ref{algo:distributed-search}}
    \State \textit{neighborlist} = Search(\textit{graph}, \textit{node}, $k$)
    \State
    \State In parallel: \Comment{Update with Algorithm~\ref{algo:update}}
    \State Update(\textit{graph}, \textit{node}, \textit{neighbors})
    \State
    \State \textit{medoid} = NearestMedoid(\textit{node}) \Comment{Shuffle}
    \State assign $\langle \textit{node}, \textit{neighborlist} \rangle$ to the compute node 
    \State corresponding to \textit{medoid}
    \State
    \State In parallel:  \Comment{Update medoids}
    \State compute new medoids
  \end{algorithmic}
\end{algorithm}

\section{Experimental evaluation}
\label{sec:evaluation}

\subsection{Datasets}

For the experimental evaluation of the algorithms, we use various datasets and similarity measures, which we explain below\footnote{Instructions to download and process the datasets can be found at \url{https://github.com/tdebatty/java-datasets}}.

\subsubsection{Synthetic dataset}

The synthetic dataset consists of points in $R^3$ which are randomly generated according to a mixture of gaussian distributions. The similarity used to build and query the graphs is the classical euclidean distance.

\subsubsection{SPAM dataset}
The SPAM dataset contains the subject of approximatively 1 million spams collected by Symantec Research Labs in 2010. Domain knowledge suggests that the most relevant similarity for building and querying the graph is Jaro-Winkler string distance. It is similar to classical Levenshtein edit distance, but Jaro-Winkler allows character substitution. Moreover, the substitution of two close characters is considered less important then the substitution of two characters that a far from each other. Also, Jaro-Winkler is not a metric distance as it does not abide by triangle inequality. 

\subsubsection{Wikipedia dataset}
The Wikipedia dataset consists in a complete dump of the static pages of \url{http://fr.wikipedia.org} made in 2008. The dataset contains a total of 2.6 million pages. To compute the similarity between pages, we first strip html markup, and use cosine similarity between 2-grams: the pages are converted into sets of 2-grams (sequences of two characters). Each page is hence represented as a vector in a multidimensional space where each dimension is a possible 2-gram. The similarity between pages is the cosine of the angle between those vectors.

\subsection{Improved Graph based Nearest Neighbor Search}

\subsubsection{Expansion parameter}

We study here the influence of the expansion parameter. We use a java implementation of iGNNS\footnote{The source code of iGNNS is available at \url{https://github.com/tdebatty/java-graphs}} and compare it to our own java implementation of GNNS. The tests are executed on a Core i7 quad core workstation equipped with 16GB of RAM memory.

\begin{figure}
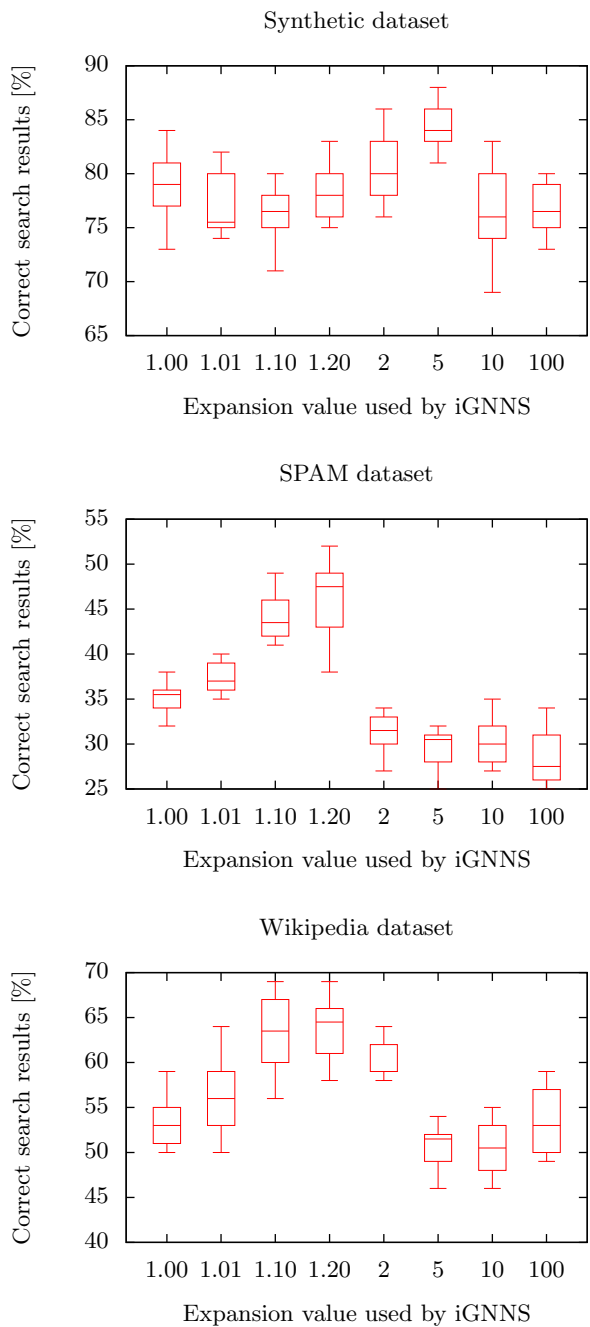

  \centering
  \input{figures/expansion-synthetic.tex}
  \input{figures/expansion-spam.tex}
  \input{figures/expansion-wiki.tex}
  \caption{Influence of iGNNS expansion parameter}
  \label{fig:expansion}
\end{figure}

For each dataset, we use a subset of given size and build the $k$-nn graph. We then perform 100 search queries using iGNNS using a fixed speedup (number of computed similarities compared to exhaustive search), and we vary the expansion parameter. For each test, we compare the results of iGNNS to the ground truth computed using exhaustive search. We repeat each experiment 10 times, using random sub-sampling validation (Monte Carlo cross-validation)\footnote{The code used to run the tests is available at \url{https://github.com/tdebatty/graph-experiments}}. The parameters used for the test are listed below, and the results can be seen in Figure~\ref{fig:expansion}.

  \vspace{3pt}
  {\setlength{\parindent}{0cm}

  \begin{tabularx}{\linewidth}{X X X}
  \hline
    Dataset   & Graph size & Speedup \\ \hline
    Synthetic & 5000       & 50       \\ 
    Spam      & 2000       & 5        \\ 
    Wikipedia & 500        & 5      \\ \hline
  \end{tabularx}
  }

As we can see, for the SPAM and Wikipedia datasets there is a clear maximum for an expansion value of $1.2$. This is not as clear for the synthetic dataset, but a value of $5$ seems to give better results. We will use these parameters for the rest of this paper. The automatic determination of the expansion parameter for new datasets is left as a future work.

\subsubsection{Comparison with GNNS}

We can now experimentally compare the improved Graph based Nearest-Neighbor Search (iGNNS) algorithm to the previously existing GNNS algorithm. In~\cite{Hajebi2011} the authors showed that, in some cases, GNNS was already more than twice faster than existing LSH or randomized kd-tree based methods, for the same quality of search.

For each dataset, we use a subset of fixed size (see Table~below) and build the $k$-nn graph. We then perform 100 search queries using GNNS and iGNNS using different speedups, and compare the result of each algorithm to the ground truth computed using exhaustive search. Once again, we repeat each experiment 10 times.

\vspace{3pt}
{\setlength{\parindent}{0cm}

\begin{tabularx}{\linewidth}{X X X}
\hline
  Dataset   & Graph size & Expansion \\ \hline
  Synthetic & 5000       & 5 \\ 
  Spam      & 2000       & 1.2 \\ 
  Wikipedia & 1000       & 1.2 \\ \hline
\end{tabularx}
}

\begin{figure}
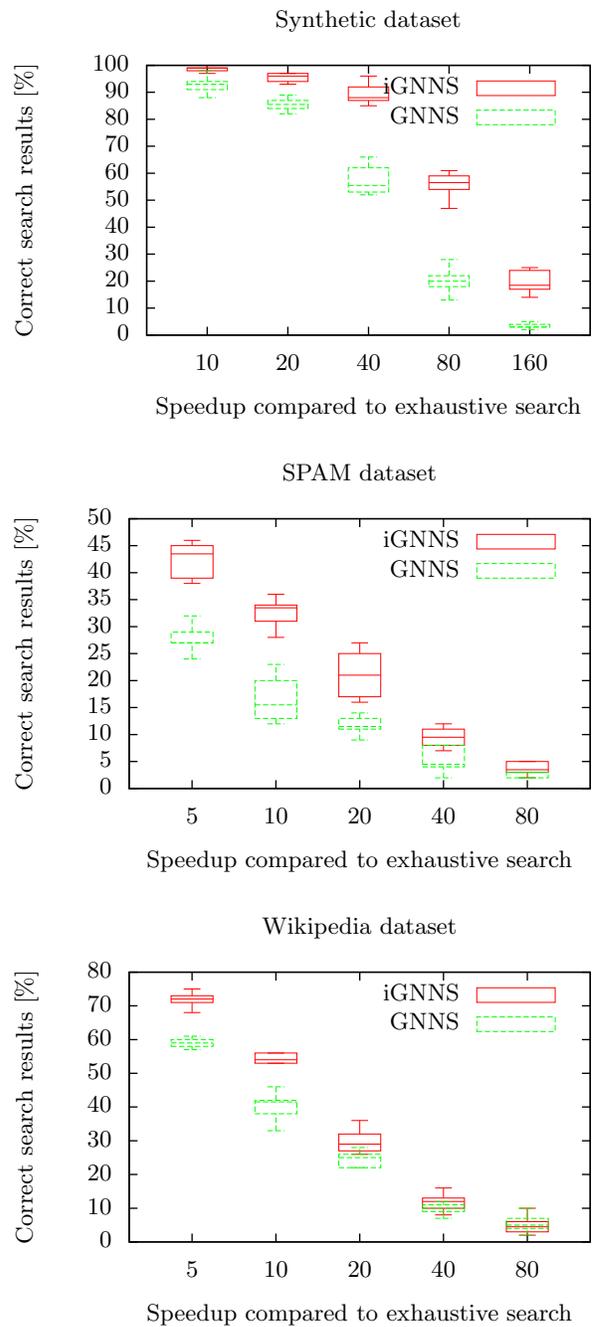

  \centering
  \input{figures/iGNNS-synthetic.tex}
  \input{figures/iGNNS-spam.tex}
  \input{figures/iGNNS-wiki.tex}
  \caption{Comparison of iGNNS and GNNS for various datasets}
  \label{fig:iGNNS}
\end{figure}

The resulting number of correct search results is shown in Figure~\ref{fig:iGNNS}. As we can see, iGNNS always outperforms GNNS, which confirms our two improvements allow iGNNS to reach the same quality of search using up to two times less similarity computations.

\subsection{Online graph building}

We evaluate here the influence of the update depth parameter used by both sequential and distributed online graph building algorithms. We build an initial $k$-nn graph of size $n$, then use the sequential online algorithm to add $n_a$ nodes to the graph, as the results are easier to interpret. We use a fixed speedup and vary the update depth from one to five.

\vspace{3pt}
{\setlength{\parindent}{0cm}

\begin{tabularx}{\linewidth}{X X X X}
\hline
  Dataset   & $n$  & $n_a$ & Speedup \\ \hline
  Synthetic & 1000 & 1000 & 20          \\ 
  Spam      & 1000 & 1000 & 10      \\ 
  Wikipedia & 1000 & 1000 & 10     \\ \hline
\end{tabularx}
}

\begin{figure}
  \centering
  \input{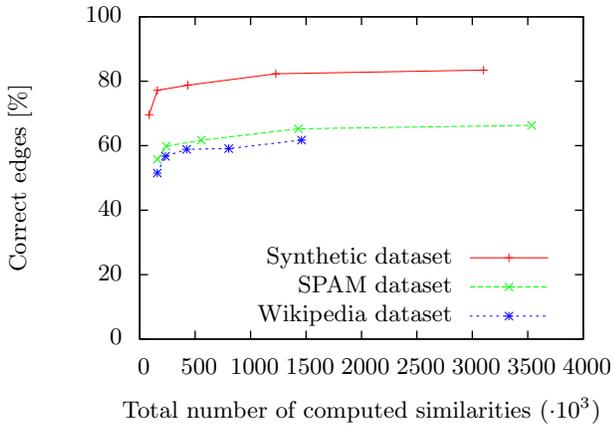}
  \caption{Influence of depth parameter for online graph building}
  \label{fig:online-depth}
\end{figure}

For each experiment, we measure the number of correct edges in the online graph, and the total number of similarities that were computed to add the $n_a$ nodes to the graph. The resulting values are presented in Figure~\ref{fig:online-depth} for the different datasets.

On the Figure, for each dataset, the first point on the left corresponds to a depth of one, while the last point of the series corresponds to a depth of five. As we can see, for every dataset there is a typical ``diminishing returns'' effect: increasing the depth above two results in an exponentially increasing number of computed similarities, while the quality of the final graph does not increase proportionally. Hence from now on we use an update depth value of two.

\subsection{Distributed search}

We can now head over to the distributed algorithms. In this paper the implementation of our distributed algorithm is designed for the spark parallel processing framework\footnote{The source code is available at \url{https://github.com/tdebatty/spark-knn-graphs}}. The experiments are run on a cluster consisting of 8 compute nodes plus one master node, each equipped with a quad-core processor and 8GB of RAM memory.

\begin{figure}
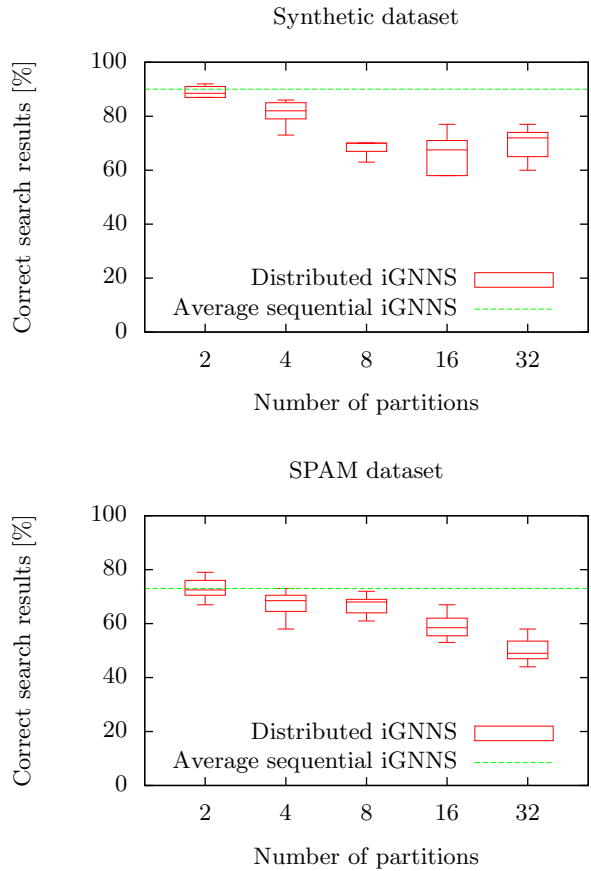

  \centering
  \input{figures/search-synthetic.tex}
  \input{figures/search-spam.tex}
  \caption{Comparison of sequential and distributed iGNNS}
  \label{fig:distributed-search}
\end{figure}

We first evaluate the distributed graph based search algorithm. Therefore, for each dataset, we build a $k$-nn graph and partition the graph using a varying number of partitions. We then perform 100 search queries using distributed iGNNS and compare the search results to the results obtained using sequential iGNNS. We repeat each experiment 10 times. The parameters used are listed below, and the results are shown on Figure~\ref{fig:distributed-search}.

\vspace{3pt}
{\setlength{\parindent}{0cm}

\begin{tabularx}{\linewidth}{X X X}
\hline
  Dataset   & Graph size  & Speedup \\ \hline
  Synthetic & 8000 & 4          \\ 
  Spam      & 4000 & 4      \\ 
\end{tabularx}
}

As we can see, using a distributed approach induces a small price in the quality of search results compared to the sequential search algorithm. As stated above, the parallelism of the algorithm is limited by the number of partitions used. One would be tempted to use a very large number of partitions. However, for the SPAM dataset, the decrease becomes really noticeable when using more than 8 partitions for a dataset of 4000 items. For the synthetic dataset, the limit is less clear but using 4 partitions seems to be the maximum. This suggests that the parallelism that can be achieved is limited by the size of the dataset: 2000 items per partition for the synthetic dataset, and 500 items per partition for the spam dataset.

\subsection{Distributed online graph building}

We now experimentally evaluate the distributed online graph building algorithm.

\subsubsection{Interval between recomputing medoids}

In the distributed online graph building algorithm, the medoids can be recomputed after a certain number of nodes have been added to the graph. We show here that this is not required for datasets for which the characteristics do not evolve over time, like our synthetic dataset.

\begin{figure}
  \centering
  \input{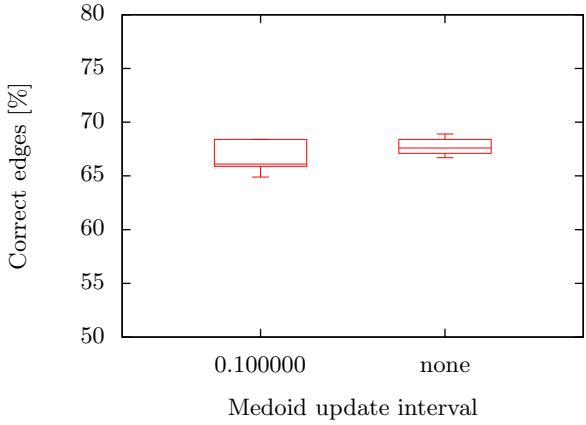}
  \caption{Number of correct edges with and without medoids update}
  \label{fig:mur}
\end{figure}

We use an initial graph of 4000 elements, then add 4000 nodes. For the first experiment we use a medoid update interval of $10\%$. Hence the medoids are first recomputed after 400 nodes have been added to the graph, then after 440 nodes have been added, and so forth. In the second case we don't update the medoids. Each experiment is repeated 10 times. The number of correct edges in the final graphs is shown in Figure~\ref{fig:mur}. As we can see, for this dataset, updating the medoids produces no noticeable difference in the final graph.

\subsubsection{Graph quality}

We now evaluate the quality of the graph produced by the online building algorithm. Therefore we build an initial $k$-nn graph using a brute force algorithm. Then we progressively add new points to the graph using the online algorithm. We regularly compare the online graph to a graph built from the same datapoints using an offline brute force algorithm. At each step we count the number of edges that are correct in the online graph $e_c$.

If the initial graph has $n$ nodes and we add $n_a$ nodes to the graph, the algorithm has to create $n_a k$ new edges. We can also expect that a number of edges from the initial graph have to be modified. Hence the total expected number of modified edges is:

$$e_m = n_a k + n k \frac{n_a}{n_a + n} $$

The expected number of unmodified edges in the final graph is simply $e_u = (n + n_a)k - e_m$. Hence the quality of the produce graph can be measured as the number of correctly modified edges divided by the expected number of edges that the algorithm should modify:

$$ Q = \frac{e_c - e_u}{e_m} $$

\begin{figure}
  \centering
  \input{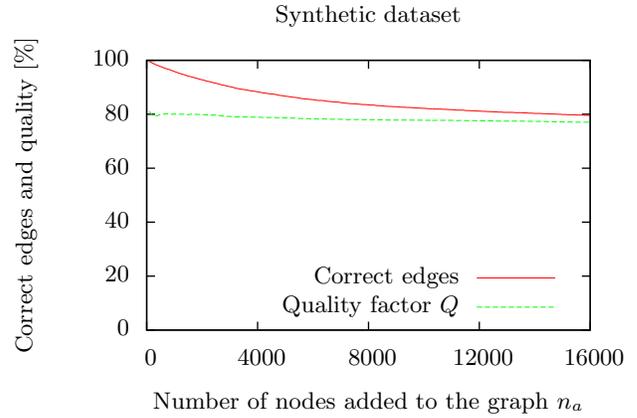}
  \input{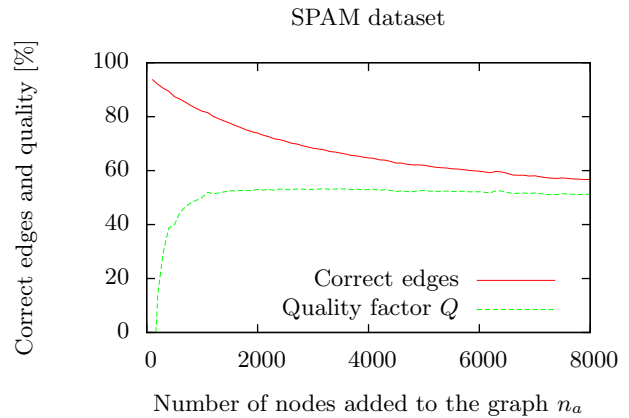}
  \caption{Quality of the graph produced by the distributed online building algorithm}
  \label{fig:online-quality}
\end{figure}

The parameters used for this experiment are listed below and the results are presented in Figure~\ref{fig:online-quality}

\vspace{3pt}
{\setlength{\parindent}{0cm}

\begin{tabularx}{\linewidth}{X X X}
  \hline
  Experiment     & quality-synthetic & quality-spam \\ \hline
  Dataset        & Synthetic        & SPAM        \\
  $n$            & 8000             & 4000        \\
  $n_a$          & 16000            & 8000        \\
  Search speedup & 4                & 4           \\
  Partitions     & 8                & 8           \\ \hline
\end{tabularx}
}

As we can see, the additive approximation induced by progressively adding new nodes to the graph has a very limited impact on the quality of the graph. It will decrease only very slowly, despite the fact that the algorithm uses a speedup of four compared to exhaustive search.

\section{Conclusions and future work}
\label{sec:conclusion}

In this paper we proposed an online approximate k-nn graph building algorithm, which is able to quickly update a k-nn graph using a flow of data points. We also proposed a distributed partitioning method based on balanced k-medoids clustering, and an improved sequential search procedure. We tested the algorithm using various datasets, and showed the distributed online graph building algorithm is able to build a graph that is highly similar to the graph produced by an
offline exhaustive algorithm, while it requires less similarity computations.

As a future work, we plan to study the auto-evaluation of the expansion and medoids update interval parameters.

\bibliography{paper}

\end{document}